\documentclass[11pt]{article}
\usepackage{hyperref}
\pdfoutput=1
\usepackage{graphicx}
\begin{document}

\title{Impinging Jet Resonant Modes at Mach 1.5}
\author{T.B. Davis and F.S. Alvi\\
\\\vspace{6pt} Department of Mechanical Engineering, \\ Florida State University, Tallahassee, FL 32306, USA}
\maketitle

%% The abstract (in this file, and that submitted as text to arXiv) should
%% include the exact phrase
%% "fluid dynamics video" or "fluid dynamics videos"

% main text

\begin{figure}[h]
\centering
\includegraphics[width = 0.5\textwidth]{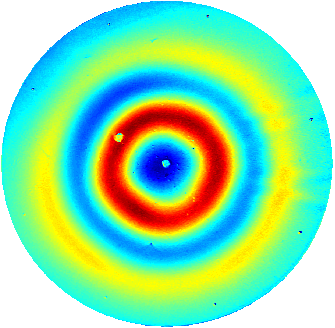}
\caption{Phase-locked PSP result showing axisymmetric mode at $h/D_{j} = 4.5$.}
\label{fig:hd4p5}
\end{figure}

\begin{figure}[h]
\centering
\includegraphics[width = 0.5\textwidth]{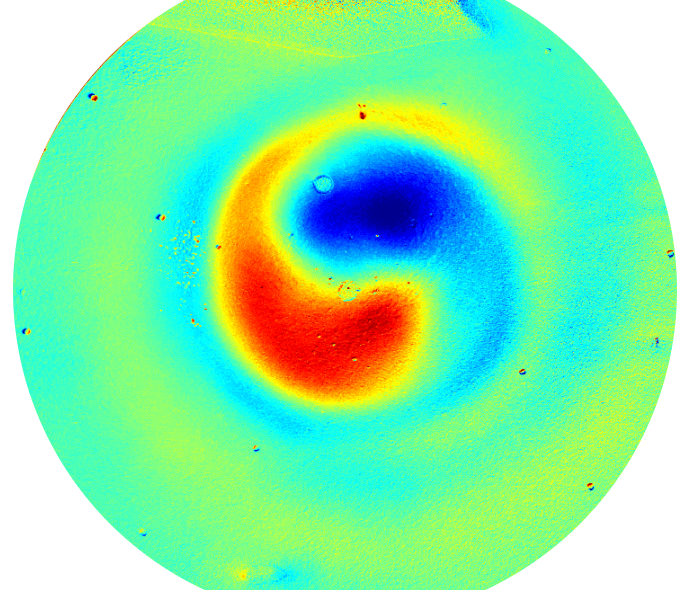}
\caption{Phase-locked PSP result showing helical mode at $h/D_{j} = 4$.}
\label{fig:hd4}
\end{figure}

High speed impinging jets have been the focus of several studies owing to their 
practical application and resonance dominated flow-field. The current study focuses on the identification and visualization of the resonant modes at certain critical impingement heights for a Mach 1.5 normally impinging jet. These modes are associated with high amplitude, discrete peaks in the power spectra and can be identified as having either axisymmetric or azimuthal modes. Their visualization is accomplished through phase-locked Schlieren imaging and fast-response pressure sensitive paint (PC-PSP) applied to the ground plane. \\
\indent 
In terms of image acquisition, a similar setup is used for both the PSP and Schlieren. An unsteady pressure transducer surface mounted on the impingement surface is used as a reference signal for the phase-locked measurements. For the PSP, an LED with a wavelength of 460 nm is used to illuminate the ground plane. The paint fluorescence ($>$ 600 nm) is then recorded with a CCD camera equipped with a long-pass filter. As the resonant frequencies are typically several kHz, the temporal resolution is limited by the camera. Due to mode switching, intermittency, and the superposition of multiple modes, it is insufficient to acquire images at the rate limited by the camera, typically $\mathcal{O}$(Hz), while still capturing the relevant dynamics. To circumvent this, the light source was triggered every cycle with minimal pulse width, with the camera shutter left open. This allows for averaging directly on the CCD and greatly improves the number of averages, minimizing the time-scales over which the averages are taken. A similar setup was used to acquire the Schlieren images. This method provides for several thousand averages taken over as many physical cycles. While such a large number of averages `washes-out' all finer details of the jet, it does allow one to distinguish different jet modes even when multiple tones are observed in the power spectrum. \\
\indent 
Results are presented in the accompanying fluid dynamics video for two nozzle to plate spacings of $h/D_{j} = 4 \mbox{ and } 4.5$. Here, $h$ is the distance from the nozzle exit to the impingement surface, and $D_{j}$ is the nozzle throat diameter. At $h/D_{j} = 4.5$, a single dominant tone is observed in the pressure spectra at $f = 6.3$ kHz. This produces a very `clean' signal for both the Schlieren and PSP, for which 8 phases are recorded. Both clearly identify the jet mode as being axisymmetric. A sample PSP image is given in Fig. \ref{fig:hd4p5}. The experimental data is supplemented with isosurfaces of fluctuating pressure from a Dynamic Mode Decomposition (DMD) of a fully 3D simulation (Uzun et al.). The fluctuating pressure is reconstructed from the DMD mode corresponding to the jet resonant frequency. The pressure isosurfaces exhibit an annular structure of alternating low/high pressure regions that are clearly associated with the axisymmetric structures observed in the Schlieren images. The results are compared with the PSP data collected on the ground plane and an excellent agreement is found.  
\\
\indent
At $h/D_{j} = 4$, multiple tones are observed in the spectra at frequencies of 4.3, 5.7 and 7.1 kHz. Data are presented at 4.3 and 7.1 kHz. The Schlieren images phase-locked at 4.3 kHz exhibit a distinct helical motion of the jet, although a significant number of averages were required to isolate the motion from the other modes. The PSP results agree, revealing a pressure footprint that decidedly resembles a `yin-yang' pattern, see Fig. \ref{fig:hd4}. The results at $f = 7.1$ kHz are less clear, possibly exhibiting a mixed mode. Closer to the nozzle exit, the jet displays a similar pulsing similar to the axisymmetric mode observed at $h/D_{j} = 4.5$. However, downstream the jet more clearly reveals a higher order helical mode. The PSP images show a similar annular structure identified with the axisymmetric mode; however, opposite phases also reveal asymmetric consistent with a helical or flapping motion. 

\indent The authors would like to thank the Florida Center for Advanced Aero-Propulsion (FCAAP), AFOSR and NSF. Also, we would like to thank Daisuke Yorita and Adam Edstrand for their help in collecting the PSP data and Ali Uzun for the simulation data provided for comparison.

\end{document}